\documentclass[aps,12pt,a4paper]{revtex4-2}
\usepackage{epsfig}
\usepackage{graphicx}
\usepackage{amsmath,amssymb,color}
\usepackage[english]{babel}
\usepackage{natbib}
\usepackage{xcolor}
\usepackage{float}
\parskip=\medskipamount
\usepackage{subfigure}

\newcommand{\eq}[1]{(\ref{#1})}
\newcommand{\fig}[1]{Fig.~\ref{#1}}

\newcommand{\be}{\begin{equation}}
\newcommand{\ee}{\end{equation}}
\newcommand{\beq}{\begin{equation}}
\newcommand{\eeq}{\end{equation}}
\newcommand\disp{\displaystyle}

\newcommand{\eps}{\varepsilon}

\begin{document}

\title{Non-algebraic first return probability of a stretched random walk near a convex boundary and its effect on adsorption}

\author{Daniil Fedotov$^{1}$, Sergei Nechaev$^{2,3}$}

\affiliation{$^1$UFR Sciences, Universit\'e Paris Saclay, 91405 Orsay Cedex, France \\
$^3$LPTMS, CNRS -- Universit\'e Paris Saclay, 91405 Orsay Cedex, France \\ $^3$Laboratory of Complex Networks, Center for Neurophysics and Neuromorphic Technologies, Moscow, Russia}


\begin{abstract}

The $N$-step random walk, elongated in the vicinity of a disc (in 2D) or a sphere (in 3D) of radius $R$, demonstrates a non-algebraic stretched exponential decay $P_N\sim \exp\left(-{\rm const}\, N^{1/3}\right)$ for the first return probability $P_N$ in the double-scaling limit $N=\frac{L}{a}\gg 1, \frac{R}{a}\gg 1$ conditioned that $\frac{L}{R}=c={\rm const}$. Stretching means that the length of the walk, $L=Na$ (where $a$ is the unit step length) satisfies the condition $L = cR$, where $c > \pi$ and under "first return" we understand the radial first arrival to a boundary. Both analytic and numerical evidences of the non-algebraic behavior of $P_N$ are provided. Considering the model of a polymer loop stretched ("inflated") by external force, we show that non-algebraic behavior of $P_N$ affects the adsorption of a polymer at the boundary of a sticky disc in 2D, manifesting in a first order localization transition.

\end{abstract}

\maketitle

\section{Introduction}
\label{sect:1}

A central quantity in the analysis of stochastic processes is the first-return probability, $P_N$, which measures the probability that a $N$-step random walker returns to its starting point for the first time on the $N$th step. This concept plays a key role in many areas of statistical physics and probability theory. One of classical examples is the distinction between recurrent and transient behavior in lattice random walks. The total return probability is given by ${\cal P} = \sum_{N=1}^\infty P_N$. The sum of first-return probabilities determines whether the walk is recurrent or transient: ${\cal P} = 1$ in 1D and 2D means that returns occur infinitely often, while ${\cal P} < 1$ in 3D and higher dimensions indicates that the walker has a finite probability of never returning. This fundamental property arises from the asymptotic behavior of $P_N$, whose slow decay in low dimensions ensures recurrence (see, for example \cite{Feller1968, Lawler2010}). The same mechanism describes localization behavior of random walks at sticky boundaries, where first-return statistics determine the long-time behavior of propagators and hence provides information about presence/absence of a spectral gap at criticality.

Another important issue in which first-return probabilities play a crucial role is diffusion-controlled chemical kinetics. When a diffusing particle reacts upon first contact with a target, the overall reaction rate is governed by the statistics of the particle’s first encounter. The survival probability for a particle not to react till some time can be written in terms of the first-return probabilities thus manifesting in a non-exponential diffusion-limited kinetics \cite{Redner2001, Hughes1995}. 

In classical settings, such as the simple symmetric random walk in a $D$-dimensional space, the first return probability, $P_N$, exhibits an algebraic decay, $P_N \sim N^{-(1+D/2)}$ for $D<2$ and $P_N \sim N^{-D/2}$ for $D>2$ (at the critical dimension $D=2$ the first return probability $P_N$ has logarithmic corrections). However, there are deviations from the algebraic form of $P_N$ that occur in more complex systems, such as, for example Griffith phase in disordered environments \cite{dotsenko}, stretched exponents in fractional Brownian motion and other types of non-Markovian dynamics \cite{bouchaud}. In particular, stretched exponential decay of the form $P_N \sim \exp(-c N^\gamma)$ with $0<\gamma<1$ is found in the "trapping problem", where the survival probability of a particle diffusing in a medium with randomly placed static traps decays as a stretched exponent due to the dominance of rare, trap-free regions. This phenomenon was first rigorously studied in the celebrated works of Donsker and Varadhan \cite{donsker} and of Balagurov and Vaks \cite{balagurov}, where a large-deviation framework were developed for such problems. Stretched exponential behavior can also arise in disorder-free systems with hierarchical or constrained geometry, such as ultrametric diffusion \cite{avetisov}, random walks on certain graphs and fractals and on percolation clusters \cite{jullien}, where the underlying structure leads to a non-algebraic probability distributions.

In contrast, stretched exponential decay in pure, disorder-free and memory-less systems is rather rare, and typically signals the presence of strong geometric constraints or long-range interactions. In our paper, we consider a simple and analytically tractable model that exhibits this kind of non-algebraic behavior: a stretched (elongated) random walk constrained to remain close to a surface of the impermeable disc (in 2D) or a sphere (in 3D). Despite the absence of any external disorder, the interplay between geometric confinement and the large deviation behavior of elongated trajectories induces a nontrivial radial path statistics, leading to a stretched exponential form of the radial first return probability to the boundary of a convex void. Our model provides an example of a "pure" system operating in the large-deviation regime, where elongation of the walk forces it to stay in a highly improbable region of a configurational space, thus leading to atypical probabilistic properties. Our goal is to demonstrate how geometry alone can fundamentally modify return statistics even in disorder-free settings.

\begin{figure}[ht]
\includegraphics[width=0.8\textwidth]{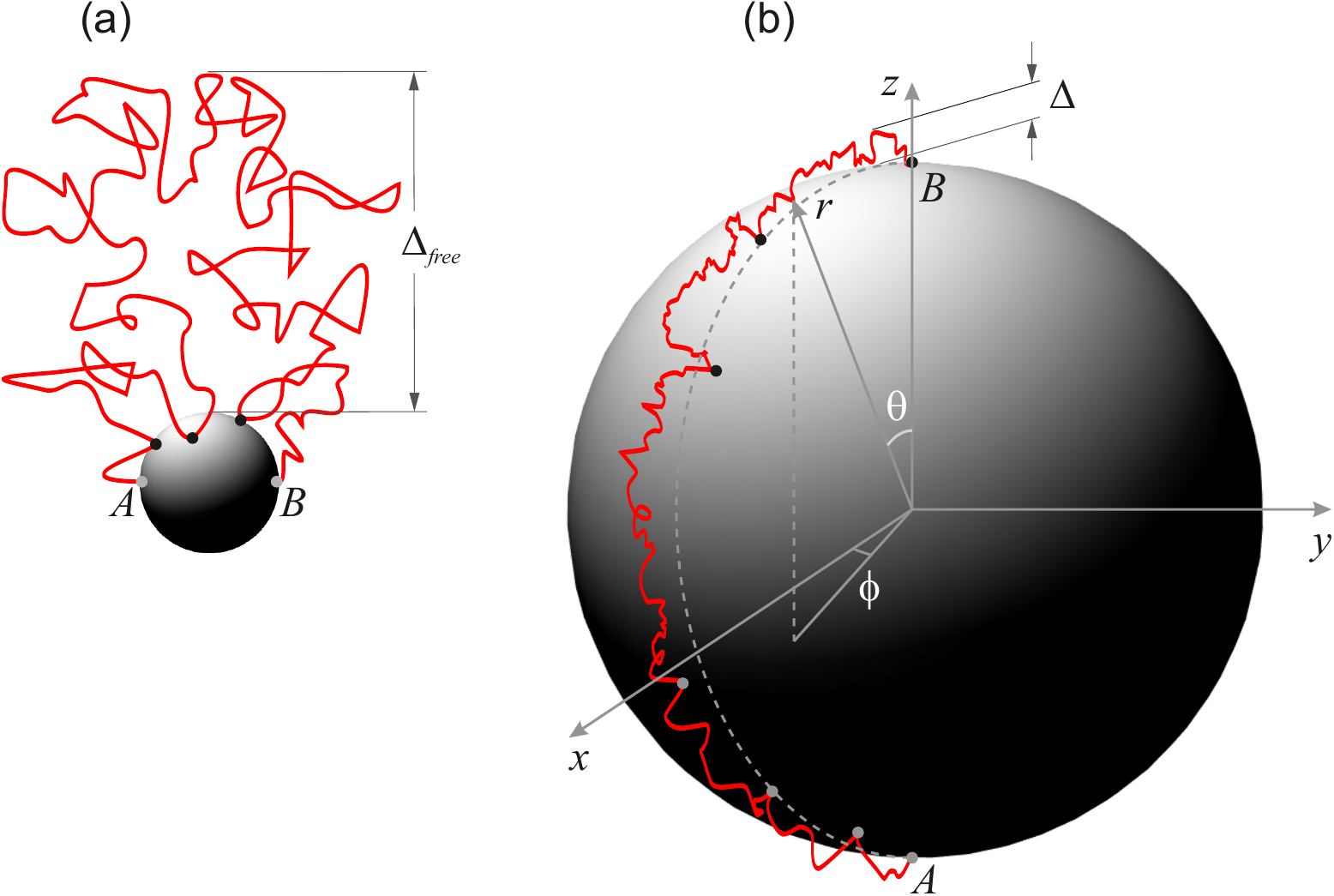}
\caption{Random walk wandering near the sphere: (a) The span of a chain is much larger than the sphere radius, $\sqrt{N}a\gg R$ -- typical (non-stretched) case; (b) The chain is elongated above the sphere: its length, $L=Na$, is proportional to the radius $R$ of a sphere, $Na = cR$ (with $c>\pi$) -- stretched case.}
\label{fig:01}
\end{figure}

Specifically, we revisit the problem of statistics of $N$-step random walk (or an ideal polymer) staying above the impermeable disc (or sphere) of radius $R$. It is typically assumed that the random walk's radial span, $\Delta_{free}(N)$, is much larger than $R$, i.e., $\Delta_{free}(N)\sim \sqrt{N}a \gg R$ (where $a$ is the unit step size). However, here we focus on a special doubly scaling regime in which the polymer is stretched above a convex, impermeable surface, i.e. when the polymer length, $L = Na$, is proportional to the radius $R$ of the convex void (a disc in $2D$ or a sphere in $3D$): $Na = cR$ (with $c$ a dimensionless numerical constant, $c \ge \pi$) for any $N \gg 1$. Schematically, these “conventional” (non-stretched) and “unconventional” (stretched) regimes in $3D$ space are depicted in \fig{fig:01}a and \fig{fig:01}b, respectively. The span of the random walk in the direction normal to the surface of a sphere (or a disc) is denoted as $\Delta$. 

Both the radial span, $\Delta(N)$, and the radial first return probability, $P_N$, of an $N$-step random walk above the convex void depend crucially on the degree of stretching (elongation) of the path. In next Section, we define precisely what we mean by stretching and by first return probability.

The "stretching regime" has been analyzed in a set of publications \cite{nech-pol,lif-tail,nech-val,nech-gr,nech-shl,meerson1,meerson2,meerson3} where it was shown that the typical span, $\Delta$, of paths above the impermeable disc in 2D as a function of $N$, follows the scaling $\Delta(N) \sim N^{\gamma} a$ with the exponent $\gamma=1/3$. Generalizing the 2D optimal fluctuation approach \cite{nech-gr} to the 3D case, we compute the scaling exponent of the span $\Delta(N)$ of the stretched path above the impermeable convex void, and provide estimates of the radial first return probability $P_N$ to the surface in 2D and 3D. Combining exact lattice path counting with extensive Monte-Carlo simulations we confirm numerically the non-algebraic stretched exponential behavior $P_N\sim \exp(-{\rm const}\, N^{1/3})$. Considering closely related model of a forcibly pre-stretched ("inflated") random loop encircling the impermeable disc, we support by more evolved analysis the preceding scaling arguments and reproduce again the non-algebraic decay of $P_N$.

The paper is arranged as follows: in Section \ref{sect:2} we provide scaling estimates of the first return probability, $P_N$, of paths elongated above the impermeable spherical surface in 3D, or the disc in 2D and compare the conjectured non-algebraic behavior with paths counting on the lattice and Monte-Carlo simulations; in Section \ref{sect:3} we derive $P_N$ for an inflated by the transversal magnetic field random loop wrapping around impermeable disc in 2D; in Section \ref{sect:4} we demonstrate the effect of non-algebraic return probability of the inflated loop on its adsorption transition at the boundary of sticky disc; finally, in Section \ref{sect:5} we summarize obtained results.

\section{Optimal fluctuations of a stretched polymer near an impermeable spherical surface}
\label{sect:2}

\begin{figure}[ht]
\includegraphics[width=0.8\textwidth]{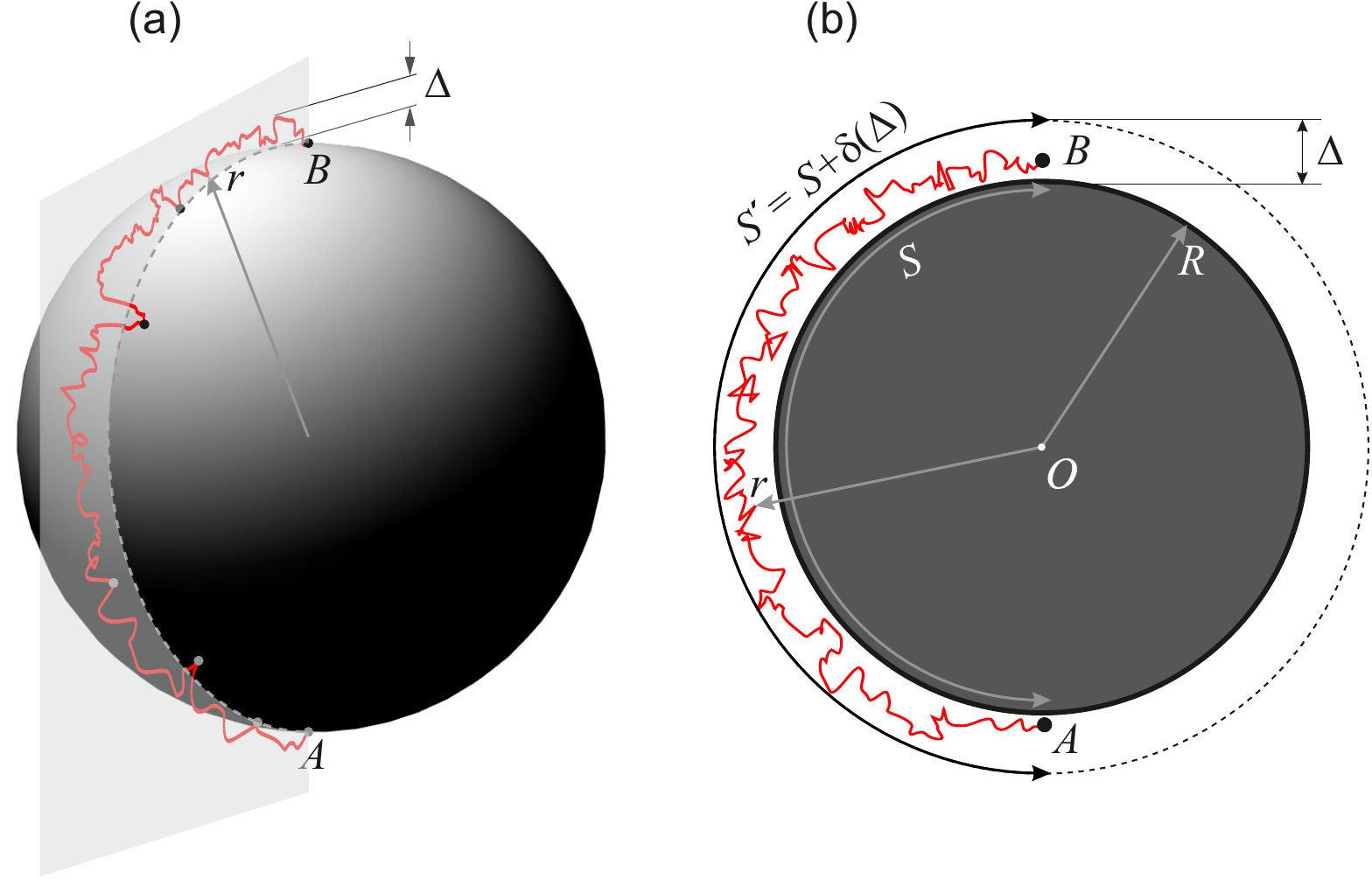}
\caption{(a) Stretched random walk wandering near the impermeable sphere in $3D$; (b) Stretched random loop wandering near the impermeable disc in $2D$.}
\label{fig:02}
\end{figure}

Schematically, the $N$-step random walk with fixed ends $A$ and $B$ stretched near an impermeable sphere of radius $R$ is shown in \fig{fig:02}a, while \fig{fig:02}b illustrates the corresponding $2D$ model. The points $A$ and $B$ are located at opposite sides of the sphere (in $3D$) or disc (in $2D$) and the overwhelming majority of paths is confined within the curved shell (in $3D$) or curved layer (in $2D$) of width $\Delta$ near the disc boundary as it is shown in \fig{fig:02}b for a $2D$ case.

The stretching refers to the ensemble of two-dimensional ideal (i.e., self-intersecting) paths in the vicinity of an impenetrable disc, subject to the following constraints:
\begin{itemize}
\item The paths do not enter the disc (the probability distribution satisfies a Dirichlet boundary condition on the disc boundary);
\item The total length $L=Na$ of each path satisfies the condition $L a = c R$, where $R$ is the disc radius, $a$ is the step length (or the lattice spacing in the discrete formulation), and $c$ is a constant.
\end{itemize}
The constraint $L a = c R$, for moderate values of $c$ ($c \gtrsim \pi$), leads to an exponential suppression of backtracking, effectively forcing the paths to be nearly directed. In Appendix I, we present results of numerical simulation demonstrating the exponential decay of the backtracking probability $U(c)$ for a two-dimensional random walk of $N = cR$ steps on a square lattice with $a = 1$, as a function of $c$ at fixed $R$. In summary, backtracking events remain possible, but their probability decreases exponentially as $c$ approaches its lower threshold $c_{\min} \sim \pi$ from above.

The term "first return probability" has two distinct meanings: a radial one (which is the primary focus of our paper) and an angular one. Under a "radial first return probability" we understand the probability for the random walk to return after $N$ steps in a circular strip of width $\delta$ in the vicinity of boundary of the sphere in $3D$, or disc in $2D$ \emph{for the first time}. To the contrary, the "angular first return probability" refers to the first arrival after $N$ steps at the specific angle $\theta$ (see \fig{fig:01}b). The angular probability has a simple exponential decay with decreasing $c$ up to some critical value $c_{min}$, i.e. when paths become more and more "directed" -- see Appendix \ref{sect:app} for discussion of this question.

In both two-- and three--dimensional cases, the problem can be formulated as solving a diffusion equation in a special large-deviation regime, where relatively short paths are constrained to stay above a convex void. If the desired asymptotic behavior of the probability distribution could be obtained solely by analyzing the solution of the diffusion equation with a Dirichlet boundary condition in the double-scaling limit $R \to \infty$, $N \to \infty$ under the constraint $N a / R = c = \mathrm{const}$, this would have provided a complete solution. However, we encountered significant technical difficulties in analyzing this limit directly. For this reason, we chose to divide the problem into two parts:
\begin{itemize}
\item $3D$ case: We employ the optimal-fluctuation approach, balancing the "elastic" and "confinement" contributions to the free energy. This yields a physically transparent picture and allows us to deduce the scaling form of the return probability.
\item $2D$ case: We solve the diffusion equation by temporarily "relaxing" the constraint $N a = c R$ and then restoring it self-consistently. Specifically, we "inflate" the random path which encircles the impermeable disc by introducing an external transverse magnetic field, and subsequently adjust a strength of this field such that the typical loop size, $R_g$, coincides with the disc radius $R$. This selfconsistent procedure leads to an explicit asymptotic expression for the return probability.
\end{itemize}
Because, in $3D$, stretched paths fluctuate primarily within the plane orthogonal to the surface, we expect the asymptotic forms in $2D$ and $3D$ to coincide. The definition of the return probabilities in $2D$ and in $3D$ are identical.

The "optimal fluctuation" arguments were originally formulated for a $2D$ system in \cite{grosb} and in a slightly different setting in \cite{lif-tail}, while later extended in \cite{nech-gr} to the generic case of fractal polymers in $2D$. The main idea proposed in \cite{grosb} and in \cite{lif-tail} is to apply the Flory-like mean-field arguments writing the free energy, $F=F_{el}+F_{conf}$, of the polymer wandering in $2D$ near a disc of radius $R$ as the sum of two contributions: (i) the "elastic" Gaussian contribution, $F_{el}$, resulting from stretching a path around a disc of radius $R$, and (ii) the "confinement" contribution, $F_{conf}$, arising from the polymer trapping in a curved layer of width $\Delta$ above the disc. Minimizing $F\equiv F(\Delta)$ with respect to $\Delta$ we find the "optimal" span of the path, $\Delta^*(N)$. 

\subsection{Elastic contribution to free energy} 

The elastic contribution, $F_{el}$, to the free energy of random walk elongated by distance $\delta\equiv \delta(\Delta)$ from the initial end-to-end distance $S$ to the final end-to-end distance $S'=S+\delta(\Delta)$ can be written as \cite{grosb, nech-gr}:
\be
\frac{F_{el}}{k_B T}=-\ln \frac{Z_{el}(N,S+\delta(\Delta))}{Z_{el}(N,S)}
\label{eq:14}
\ee
where the partition function of a Gaussian polymer in a $3D$ space is
\be
Z_{el}(N,S) = \mu^N \left(\frac{3}{2\pi N a^2} \right)^{3/2} \exp\left(-\frac{3S^2}{2Na^2} \right)
\label{eq:15}
\ee
In \eq{eq:15} $\mu$ is a geometry-dependent nonuniversal parameter. Substituting \eq{eq:15} into \eq{eq:14} and expanding  the obtained expression up the first non-vanishing terms in $0<\tfrac{\delta(\Delta)}{S}\ll 1$, we arrive at the following simple expression:
\be
\frac{F_{el}}{k_B T} = \frac{3S\delta(\Delta)}{Na^2} + \mathrm{const}
\label{eq:16}
\ee
From the geometry of the system shown in \fig{fig:02}b we immediately conclude that
\be
\frac{S}{R} = \frac{S+\delta(\Delta)}{R+\Delta} \quad \Rightarrow \quad \delta(\Delta) = \frac{S}{R}\Delta 
\label{eq:17}
\ee
which allows us to write $F_{el}$ as
\be
\frac{F_{el}}{k_B T} = \frac{3S^2 \Delta}{R Na^2}
\label{eq:18}
\ee
where we dropped off the irrelevant constant. In what follows we assume that $S=\pi R$ what brings \eq{eq:18} into
\be
\frac{F_{el}}{k_B T} = \frac{3\pi^2 R \Delta}{Na^2}
\label{eq:19}
\ee

\subsection{Contribution to free energy from a chain confinement} 

Within the Flory-like approach the elastic part of the free energy, $F_{el}/(k_B T)$ (see \eq{eq:19}), coming from the elongation of the Gaussian chain, should be balanced with the "confinement" contribution, $F_{conf}/(k_B T)$, due to the random walk localization in a narrow shell in a $3D$ space of width $\Delta$ where $\Delta/(Na)\ll 1$. This confinement part can be written as $F_{conf}/(k_B T) = - \ln {\cal Z}(\Delta, N)$ where ${\cal Z}(\Delta,N)$ is the partition function of the $N$-step random walk in the shell of width $\Delta$. Up to the normalization factor, ${\cal Z}(\Delta,N)$ coincides with the Green function, $G({\bf R}_A, {\bf R}_B, N)$ of an $N$-step random with fixed ends in a narrow shell between two spheres of radii $R$ and $R+\Delta$ at which the Dirichlet boundary conditions are imposed. The function $G({\bf R}_A, {\bf R}_B, N)$ in the continuum limit can be estimated by solving the following boundary problem:

\be
\begin{cases}
\disp \frac{\partial}{\partial N} G({\bf r},N) = \frac{3a^2}{2} \nabla^2 G({\bf r},N)  \medskip \\
G(|{\bf r}|=R,N)=G(|{\bf r}|=R+\Delta,N)=0 \medskip \\
G({\bf r},N=0) = \delta({\bf r}-{\bf R}_A)
\end{cases}
\label{eq:20}
\ee
where the radial distance $r$ is measured from the center $O$ of the sphere, and $\nabla^2$ is the Laplace operator in spherical coordinates $(r,\phi,\theta)$:
\be
\nabla^2 = \frac{1}{r^2}\frac{\partial}{\partial r}\left(r^2 \frac{\partial}{\partial r}\right) + \frac{1}{r^2 \sin\theta}\frac{\partial}{\partial \theta}\left(\sin\theta \frac{\partial}{\partial \theta} \right) + \frac{1}{r^2 \sin^2\theta} \frac{\partial^2}{\partial \phi^2}
\label{eq:21}
\ee
The solution to \eq{eq:20}--\eq{eq:21} can be written as a bilinear expansion
\be
G({\bf R}_A,{\bf R}_B, N) = \sum_{k=1}^{\infty} e^{-\frac{3a^2}{2} \lambda_k N} \Psi_k({\bf R}_A) \Psi^*_k({\bf R}_B)
\label{eq:22}
\ee
where $\lambda_k$ and $\Psi_k({\bf R}_A)$ are the eigenvalues and eigenfunctions of the random walk in a spherical shell. The explicit expression of the Green function $G(r,\theta,\phi,N)$ with the Dirichlet boundary conditions at bounding spheres is:
\begin{multline}
G(r,\theta,\phi,N) = \sum_{\{m,n\}=0}^{\infty}\sum_{k=0}^n e^{-\tfrac{3a^2}{2}\lambda_m^{(n)} N} \frac{Z_{n+1/2}\left(\sqrt{\lambda_m^{(n)}}r\right)}{\sqrt{r}} \times \\ P_{n,k}(\cos \theta)\left(A_{m,n,k} \cos k\phi + B_{m,n,k} \sin k\phi \right) \hspace{1cm} 
\label{eq:23}
\end{multline}
where
\begin{multline}
Z_{n+1/2}\left(\sqrt{\lambda_m^{(n)}} r\right) =J_{n+1/2}\left(\sqrt{\lambda_m^{(n)}} R\right)Y_{n+1/2}\left(\sqrt{\lambda_m^{(n)}} r\right)- \\ Y_{n+1/2}\left(\sqrt{\lambda_m^{(n)}} R\right) J_{n+1/2}\left(\sqrt{\lambda_m^{(n)}} r\right)
\label{eq:24}
\end{multline}
and $J_{n+1/2}(...)$ and $Y_{n+1/2}(...)$ are correspondingly Bessel and Neuman functions. The coefficients $A_{m,n,k}$ and $B_{m,n,k}$ have rather cumbersome expressions and since in what follows we do not need them, they are not reproduced here. The eigenvalues $\lambda_m^{(n)}$ are determined by the condition of nullifying the function $Z_{n+1/2}(\lambda_m^{(n)}r)$ at the external bounding sphere of the radius $r=R+\Delta$:
\be
Z_{n+1/2}\left(\sqrt{\lambda_m^{(n)}} (R+\Delta)\right)=0
\label{eq:25}
\ee
At $N\gg 1$ the free energy of the polymer confined between two concentric spheres is determined by the contribution of the term $\lambda_m^{(n)}$ with $m=n=0$ in \eq{eq:23} (the so-called "ground state dominance"). In the case $m=n=0$ Eq. \eq{eq:25} essentially simplifies giving 
\be
Z_{1/2}=\frac{2\sin \left(\sqrt{\lambda_0^{(0)}} \Delta\right)}{\pi \sqrt{\lambda_0^{(0)} R(R+\Delta)}} 
\label{eq:26}
\ee
which together with \eq{eq:25} gives us $\lambda_0^{(0)} = \pi^2/\Delta^2$. Thus, the contribution to the free energy, $F_{conf}$, of a Gaussian chain confined in a narrow spherical shell in $3D$ has the following expression
\be
\frac{F_{conf}}{k_B T} \approx \frac{3a^2N}{2}\lambda_0^{(0)} = \frac{3\pi^2 Na^2}{2\Delta^2} 
\label{eq:27}
\ee

\subsection{Span and first return probability: scaling derivation and numerical verification}

\subsubsection{Span of stretched paths}

The free energy, $T$, of a Gaussian polymer stretched within a spherical shell located between two concentric spheres of radii $R$ and $R+\Delta$ is expressed as the sum of two terms, given by \eq{eq:19} and \eq{eq:27}:
\be
\frac{F}{k_BT} = \frac{3\pi^2 R \Delta}{Na^2} + \frac{3\pi^2 Na^2}{2\Delta^2}
\label{eq:28}
\ee
Instanton-like approach suggests minimizing $F/(k_BT)$ in \eq{eq:28} with respect to $\Delta$. Thus we get the optimal width, $\Delta^*(N,R)$, of the spherical shell which fixes the typical span of the stretched path above an impermeable sphere of radius $R$:
\be
\Delta^*(N,R) = \left(\frac{N^2a^4}{R}\right)^{1/3}
\label{eq:29}
\ee
In the situation of extreme stretching, $R = Na/c$, of the path above the sphere in $3D$, we get the following scaling dependence for the span of the stretched path:
\be
\Delta^*(N) = c^{1/3}a N^{1/3}
\label{eq:30}
\ee

\subsubsection{Non-algebraic scaling of first return probability for stretched paths}

Scaling arguments for estimating the radial first return probability, $P_N$, to the boundary of an impermeable convex void in two-- and three--dimensional space are based on a striking analogy with the "optimal fluctuation" approach used to evaluate the $N$-step survival probability, $S_N$, of a one-dimensional random walk in an array of randomly placed traps with a Poisson distribution, known as the Balagurov-Vaks (BV) problem \cite{balagurov}. Recall that $S_N$ is the probability that the random walk avoids all traps up to time $N$, i.e. $S_N$ is the probability that the walk first encounters a trap precisely at time $N$. The optimal fluctuation approach to this trapping problem involves balancing two contributions to the free energy: one arising from the confinement of the $N$-step path within an interval of size $\Delta$, and the other---from the probability of a spontaneously occurring trap-free interval of size $\Delta$ in the Poisson ensemble characterized by the mean concentration, $\lambda$, of traps per unit length. The corresponding survival probability, $S_N$, has the well-known asymptotic form, $S_N \sim e^{-\lambda^{2/3} N^{1/3}}$, at $N\gg 1$. 

Comparing the first return problem of stretched random walk above the sphere (for the random walk stretched above the disc in $2D$ the scaling dependence is the same), we see that the confinement term  in our model and in Balagurov-Vaks one have the same meaning, though might differ by numerical constants due to the difference between curved and flat geometries. Surprisingly, the elastic contribution to the free energy of path stretched above the convex void also matches at extreme stretching ($Na=cR$) the entropic contribution of the spontaneous creation the trap-free regions. Relying of this analogy, one sets a correspondence between the constants $c$ and $\lambda$ in a stretching and a trapping problems: $c \leftrightarrow \lambda^{-1}$. 

Using \eq{eq:28} and \eq{eq:29} we estimate the "survival-like" first-return probability $P_N$ of a stretched path with the exponential accuracy as follows:
\be
P_N \sim \exp\left(-\frac{F}{k_BT}\right) = \left.\exp\left(-\frac{3\pi^2 R \Delta}{Na^2} - \frac{3\pi^2 Na^2}{2\Delta^2}\right)\right|_{\Delta=\Delta^*(R,N)} = \exp\left(-\frac{9\pi^2 R^{2/3}}{2a^{2/3}N^{1/3}}\right)
\label{eq:31}
\ee
which in the double-scaling limit $\frac{N}{a} \gg 1$, $\frac{R}{a} \gg 1$ under the constraint $\frac{Na}{R} = c = \mathrm{const}$ provides the following stretched exponent behavior:
\be
P_N \sim \left.\exp\left(-\frac{9\pi^2 R^{2/3}}{2a^{2/3}N^{1/3}}\right)\right|_{R=Na/c} = \exp\left(-\frac{9\pi^2 N^{1/3}}{2c^{2/3}}\right)
\label{eq:31a}
\ee

The following two facts regarding Eqs. \eq{eq:31} and \eq{eq:31a} should be admitted: 
\begin{itemize}
\item[(i)] We cannot rely on the numerical value of the coefficient in the exponent in \eq{eq:31} -- only the scaling behavior is meaningful. Therefore, we verify the dependence on $N$ and on $c$ in \eq{eq:31} in numerical simulations.
\item[(ii)] Simulations are carried out in $2D$ (rather than in $3D$), as three-dimensional simulations are significantly more challenging. Working in $2D$ is justified because chains stretched around a spherical surface in $3D$ space tend to fluctuate primarily within the plane orthogonal to the sphere. So, we expect that scaling dependence of first return probability on $N$ and on $c$ does not depend on the space dimensionality.
\end{itemize}

\subsubsection{Numerical computation of span $\Delta$ and first return probability $P_N$ in 2D}

\begin{figure}[ht]
\includegraphics[width=0.98\textwidth]{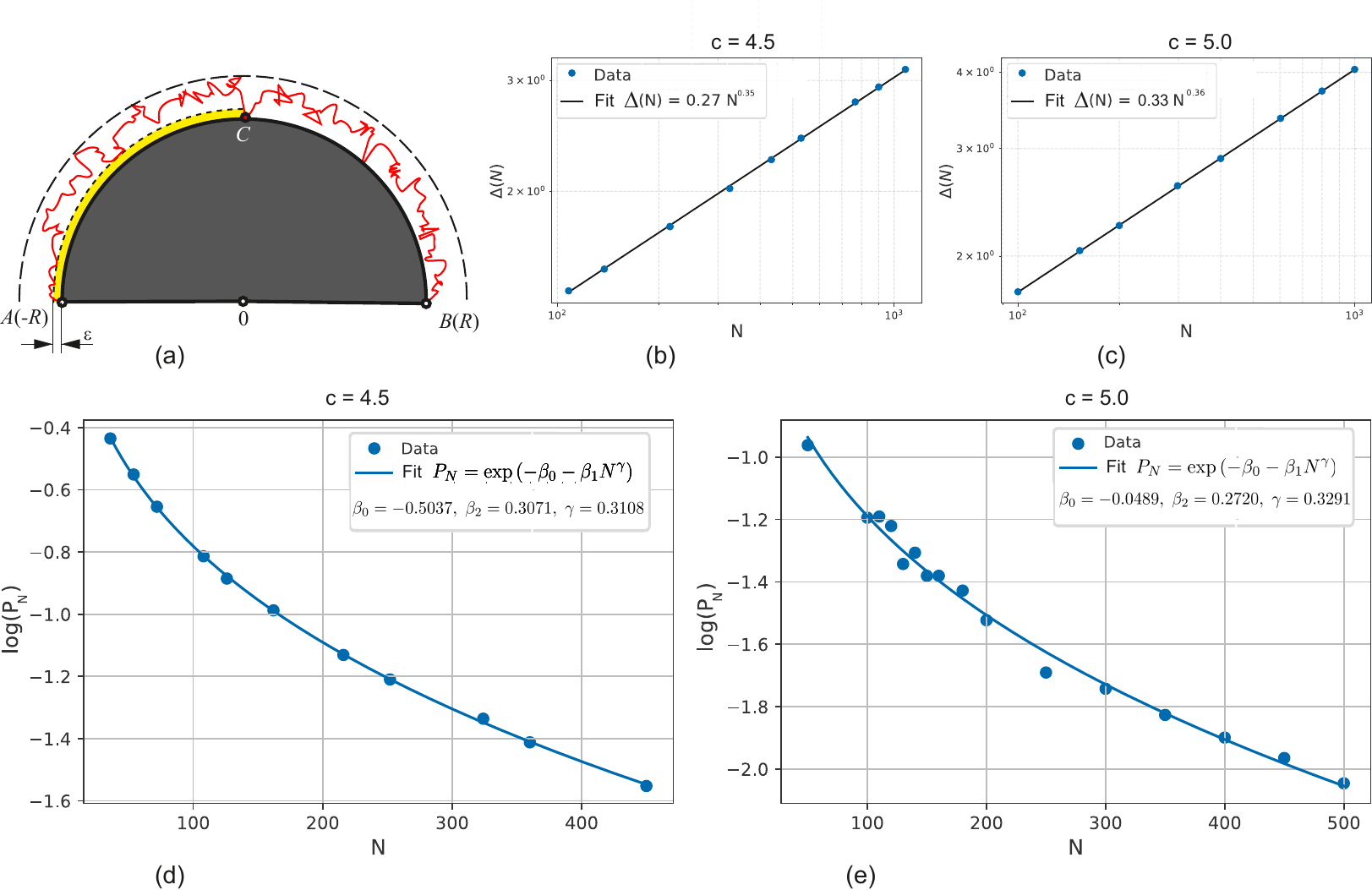}
\caption{(a) A typical configuration of a stretched random path starting at $A(-R, 0)$ and first returning to the boundary of the semicircle at the point $C$. The yellow area indicates a narrow $eps$-strip inaccessible for the random walk; (b)-(c) Log-log plot of the average span of stretched paths as a function of $R$ for $c=4.5$ (b) and $c=5$ (c); (d)-(e) Log-normal plot of the fraction of trajectories for which the random walk returns to the boundary for the first time on the step $N/2$, out of all $N$-step stretched paths with $N$ ranging from 50 to 600 for $c=4.5$ (d) and $c=5$ (e). \footnote{Source data for the figure can be found at~\cite{fedotov2025}.}}
\label{fig:03}
\end{figure}

To verify numerically expressions \eq{eq:30} and \eq{eq:31}, we follow the protocol proposed in \cite{nech-pol} for computing the typical deviation from a semicircular boundary. It is noteworthy that we use an exact lattice path-counting (but not a Monte-Carlo method), therefore confidence intervals cannot be provided. Specifically, we fix a value $R/a$ (in what follows the lattice spacing $a$ is set to unity) and enumerate all stretched bridges of $N$ steps such that $N=cR$ for a fixed value of $c$. Bridges start at a point $A$ and end at a point $B$. Computing fractions $p(\Delta)$ of paths reaching points $\Delta = 0,1,2,..$ above the tip $C$ of a disc of a given $R$ --- see \fig{fig:03}a --- we find the averaged value $\Delta^*$ as $\Delta^* =\sum_{\Delta=0}^{\Delta_{max}} \Delta\, p(\Delta)$. Repeating computations for a set of values $R$ in the interval $R=[20, 200]$ we find the dependence $\Delta^*(N)$ which fits very well \eq{eq:29} --- see \fig{fig:03}(b)-(c). Evaluations are performed for two values $c=4.5$ and $c=5$.

To compute the radial first return probability in $2D$ we proceed as follows: we fix a narrow circular strip of width $\eps$ shown in yellow in \fig{fig:03}a inaccessible for paths, and compute the number of trajectories starting from $A$ and reaching the red dot in \fig{fig:03}a which is located at the boundary of a disc (i.e. has a radial coordinate $\Delta=0$) after $N/2$ steps. The presence of a strip of width $\eps$ (in the simulations $\eps$ coincides with the lattice spacing) insures that paths do not touch the boundary until the step $N/2$ above the point $C$ of the disc. After $N$ steps the path reaches the point $B$, but the region form the point $C$ (red dot) in \fig{fig:03}a till the point $B$ is not interesting for us since after the first return at the point $C$, the ensemble of paths factorizes and the parts from $A$ to $B$ and from $B$ to $C$ are independent. The nonalgebraic scaling behavior is independent of the cutoff $\eps$ for trajectories of length $\tau$, when $\tau \gg \eps^2$. In the opposite regime, i.e., when the considered path length $\tau$ is small compared to $\eps^2$, the part of the void behind the path is effectively flat, and the first-return probability scales as in the standard $1D$ case, i.e., $\sim \tau^{-3/2}$. Thus, the observed nonalgebraic decay is a consequence of the curvature, while "stretching" determines the scale at which this behavior becomes apparent.

Using the least-square method, we numerically check the theoretical scaling laws for $\Delta(N)$ and $P_N$. For $\Delta(N)$, we have reproduced results of \cite{nech-val} in \fig{fig:03}(b)-(c), while for $P_N$ we approximate the conjectured scaling dependence by the trial fit $\exp\left(-\beta_0-\beta_1 N^{\gamma}\right)$ with three adjustable parameters, $\beta_0, \beta_1$, and $\gamma$ --- as shown in \fig{fig:03}(d)-(e). We got the scaling exponent $\gamma \in [0.31, 0.33]$ for $c=4.5$ and $c=5$. The resulting statistical characteristics are: $R^{2} = 0.999$, and a $p$-value for $\gamma$ is of order of  $\ll 10^{-3}$.  This leads us to conclude that the numerics agrees very well with the theoretically predicted functional dependence \eq{eq:31}. 

\section{Statistics of inflated Brownian paths above an impermeable disc in 2D} 
\label{sect:3}

We compute the first-return probability of stretched paths above a disc with Dirichlet boundary conditions, generalizing the line of reasoning proposed in \cite{nech-val}. At first glance, the problem appears simple, as it reduces to solving the diffusion equation in polar coordinates. However, we are specifically interested in the double-scaling regime where $N \to \infty$ and $R \to \infty$ under the constraint $Na/R = c = \mathrm{const}$. This regime corresponds to a highly improbable large-deviation limit and obtaining an analytic solution in this limit is rather challenging. To proceed, we temporarily relax the constraint $Na = cR$ and treat $N$ and $R$ as independent parameters. To compensate stretching, we “inflate” the Brownian loop surrounding the disc by introducing an external transverse constant magnetic field, whose strength is then determined self-consistently by requiring the loop radius $R_g$ to coincide with the disc radius $R$. Solving the resulting boundary-value problem yields the desired probability distribution.

Consider an ensemble of inflated two-dimensional $N$-step random loops leaning on an impermeable disc of radius $R_d$ in the Euclidean plane. Starting from a closed two-dimensional random walk of length $L = Na$, we inflate its algebraic area, $A$, until it reaches the area of a circle, $A_c = \pi R_c^2$, where $R_c = c R_d$ and $c \geq 1$. The area $A_c$ is controlled via a Lagrange multiplier, $q$, allowing the Brownian loops to fluctuate within a large deviation regime centered around $A_c$, which is atypical for non-inflated random walks. By placing an impermeable disc of radius $R_d$ inside the inflated loop, we study the corresponding Dirichlet boundary problem for random paths that stay above the disc in this large deviation regime.

The distribution function of random walks with the constraint on the algebraic area $A_d$ can be written in terms of a path integral (see, for example \cite{khandekar}):
\be
P(r,N) = \frac{1}{2\pi}\int_{-\infty}^{\infty}dq e^{-iq A_d}\int{\cal D}\{\mathbf{r}(t)\} e^{-\int_0^N L(\mathbf{r}, \dot{\mathbf{r}},t) dt}
\label{eq:c01}
\ee
where $L$ is the Lagrangian of a 2D random walk enclosing some algebraic area is as follows:
\be 
L(\mathbf{r}, \dot{\mathbf{r}},t) = \frac{1}{a^2} \left(\dot{x}^2+\dot{y}^2\right)- \frac{i q }{2}\left(x \dot{y}-y\dot{x} \right) =
\frac{1}{a^2} \left(\frac{\partial \mathbf{r}}{\partial t}\right)^2+ \frac{i q}{2} \mathbf{A}(\mathbf{r})\frac{\partial \mathbf{r}}{\partial t}
\label{eq:c02}
\ee
and the vector potential $\mathbf{A}(\mathbf{r})$ of the constant magnetic field $\mathbf{B}$ reads:
\be
\mathbf{A}(\mathbf{r}(t)) = \mathbf{B} \times \mathbf{r}(t), \quad \mathbf{B} = (0,0,B) 
\label{eq:c03}
\ee
The dimensionless "time" $t$ is changing in the interval $[0,N]$. Instead of integrating over $q$ we consider $q$ as a Lagrange multiplier and define it from the requirement that the random walk fluctuates around the classical Larmor orbit of radius $R_c=2\pi N$. Just this condition means the consideration of pre-stretched loops pushed to the atypical large deviation regime of the distribution function. To find the radius of the orbit we minimize the action  $S = \int_0^N L(\mathbf{r}, \dot{\mathbf{r}},t) dt$ with respect to the variations of paths and obtain the classical Euler-Lagrange equations whose solution are circles and can be written in the parametric form:
\be
\begin{cases}
x(t) = R \cos(\omega t) \\ y(t) = R \sin(\omega t)
\end{cases}; \quad \omega = \frac{a^2|q| B}{2}; \quad R = \frac{2}{a^2 |q| B}
\label{eq:c04}
\ee
Selection of $B$ is arbitrary, so we can chose $B=\frac{2}{a}$. In order to make the radius of Larmor orbit equal to $R_c$, we take the value of $q$ as:
\be
q = \frac{1}{a R_c}
\label{eq:c05}
\ee
The distribution function $P(r,N)$ we compute using the canonical formalism of derivation the non-stationary Schr\"odinger-like equation (see for example, \cite{edwards}). Having Lagrangian $L$ we construct the Hamiltonian $H$ 
\be
H = \frac{a^2}{4}\left(\hat{p}+ iq A\right)^2; \quad \hat{p} = \nabla_{r,\varphi} 
\label{eq:c06}
\ee
Considering $H$ as an operator acting in polar coordinates $(r,\varphi)$ on the function $P(r,N)$ we get the diffusion-like equation
\be
\frac{a^2}{4}\left(\frac{1}{r}\frac{\partial}{\partial r}\left(r \frac{\partial}{\partial r}\right) + \frac{1}{r^2}\frac{\partial^2}{\partial\varphi^2} + 2iq \frac{\partial}{\partial \varphi} - q^2 r^2\right)P(r,\varphi,t)=\frac{\partial}{\partial t}P(r,\varphi,t)
\label{eq:c07}
\ee
Recall that we have an ensemble of inflated (stretched) magnetic Brownian loops with the charge $q=\frac{1}{aR_c}$ encircling the impermeable disc of radius of $R_d$. We solve Eq. \eq{eq:c07} with Dirichlet boundary condition at the disc boundary by separating variables. Writing $P(r,\varphi,t)=e^{-\lambda t} W(r)e^{i m \varphi}$ and making use of substitution $W(r) = U(r)r^{-1/2}$ we arrive at the following set of equations for the function $U(r)$:
\be
\begin{cases}
\disp \frac{d^2U(r)}{dr^2}- \left(q^2r^2-\frac{1}{4r^2}-\frac{4}{a^2}\lambda + 2qm \right) U(r) = 0 \medskip \\
\disp U(r=R_d,t) = U(r\to\infty,t)=0
\end{cases}
\label{eq:c08}
\ee
Analyzing \eq{eq:c08} in the vicinity of the disc boundary, i.e. when $r = R_d+\rho$, where $0<\frac{\rho}{R_d}\ll 1$, we get:
\be
\frac{d^2U(\rho)}{d \rho^2}- \alpha_1 U(\rho) - \alpha_2 \rho\, U(\rho) = 0
\label{eq:c09}
\ee
where
\be
\alpha_1=q^2R_d^2-\frac{1}{4R_d^2}-\frac{4}{a^2}\lambda+2q m , \qquad \alpha_2=2q^2R_d+\frac{1}{2R_d^3}
\label{eq:c10}
\ee
The eigenfunctions of \eq{eq:c09} that satisfy zero boundary conditions are:
\be
U_j(\rho)\propto \mathrm{Ai}\big(\alpha_2^{1/3} \rho+\overbrace{\alpha_1 \alpha_2^{-2/3}}^{a_j}\big)
\label{eq:c11}
\ee
where $\mathrm{Ai}(x)=\frac{1}{\pi}\int\limits_0^\infty \cos\left(\frac{t^3}{3}+x t\right)dt$ is the Airy function and  $a_{j}<0$ are zeros of the Airy function, i.e. the solutions of $\mathrm{Ai}(a_{j})=0$ ($j=1,2,...$), where $a_1\approx -2.338$. 

Combining the condition $q=\frac{1}{a R_c}$ with the large-$R$ limit we get:
\be
\begin{cases}
\disp \alpha_2\approx 2q^2 R_d=\frac{2 R_d}{a^2 R_c^2}; \\ \disp \lambda_{jm} \approx \frac{a^2}{4}\left(q^2 R_d^2 +2 q m + |a_j|(2q^2R_d)^{2/3}\right) = \frac{1}{4}\left(\frac{R_d}{R_c} \right)^2 + \frac{m a}{2R_c} + \frac{|a_j|}{2^{4/3}}\left(\frac{aR_d}{R_c^2} \right)^{2/3}
\end{cases}
\label{eq:c12}
\ee
Using the relation $R_c = c R_d$ and taking into account that $N a =2\pi R_c$, we rewrite \eq{eq:c12} as follows:
\be
\begin{cases}
\disp \alpha_2\approx \frac{1}{\pi a^3 c N}; \\ \medskip \disp \lambda_{jm} \approx \frac{1}{4c^2}  + \frac{|a_j|}{4 (\pi c N)^{2/3}}+ \frac{m a}{4\pi N}
\end{cases}
\label{eq:c13}
\ee
The general solution to the eigenvalue problem in the presence of an impermeable disc of radius $R_d$ can be written in the form of a bilinear expansion over eigenfunctions $U_j$ with eigenvalues $\lambda_j$: 
\be
P(\rho,\rho',N)=\sum\limits_{j=1}^{\infty}\sum_{m=1}^{\infty} \frac{C_j}{R_d} e^{-\lambda_{jm} N} U_j(\rho) U_j^*(\rho)
\label{eq:c13a}
\ee
where $(...)^*$ means complex conjugation. Substituting the explicit expressions of eigenfunctions \eq{eq:c12} into \eq{eq:c13} and supposing that $m=1$ (i.e, a path makes a single turn around a disc), we arrive at the expression valid up to the first finite-size exponential correction in the large-$N$ limit:
\be
P(\rho,\rho',N)=\sum\limits_{j=1}^\infty \frac{C_j}{R_d} e^{-\frac{N}{4c^2} - \frac{|a_j|N^{1/3}}{4 (\pi c)^{2/3}}} \mathrm{Ai}\left(\frac{2^{1/3}\rho}{a(\pi c N)^{1/3}}-|a_j|\right)\mathrm{Ai}\left(\frac{2^{1/3}\rho'}{a(\pi c N)^{1/3}}-|a_j|\right)
\label{eq:c14}
\ee
where the constants $C_j$ are chosen to fulfill the initial and normalization conditions.

Equation \eq{eq:c14} allows to arrive at the following conclusions:
\begin{itemize}
\item Typical span $\Delta$ of a random walk of length $L=Na$ wandering around a disc of a radius $R_d$ in a stretched regime $N a = \pi R_c = \pi c R_d$ is:
\be
\Delta \approx |a_1| \left(\frac{\pi c}{2}\right)^{1/3} a N^{1/3}
\label{eq:c14a}
\ee
\item The distribution function $P(N)$ in the "ground state dominance" ($N\gg 1$) has the following scaling behavior up to the first finite-size exponential correction:
\be
P(N) \approx e^{-\frac{N}{4c^2} - \frac{|a_1|N^{1/3}}{4 (\pi c)^{2/3}}} 
\label{eq:c15}
\ee
\end{itemize}
As we see, the scaling dependencies on $N$ and $c$ in \eq{eq:c14a}-\eq{eq:c15} are consistent with heuristic arguments provided in Section \ref{sect:2}. However \eq{eq:c15} contains an extra linear dependence on $N$ in the exponent. The presence of this additional factor reflects the difference between the models considered here and those in Section \ref{sect:2}. Namely, in previous Section, the full ensemble of paths was considered, and stretching emerged only as a result of taking the double-scaling limit $N\gg 1, R\gg 1$ conditioned that $\frac{Na}{R}=c=\mathrm{const}$ in the final expression. In the present Section, the trajectories were "pre-inflated" by a magnetic field chosen such that it forces paths' stretching. This condition can be interpreted as a confinement condition imposed on the span of fluctuations in the paths' ensemble, and it leads to an additional exponential factor, typical for fluctuations of paths in a bounding box.

\section{Adsorption of pre-stretched random paths at the boundary of a sticky disc}
\label{sect:4}

The problem of polymer adsorption despite of a long history \cite{degennes, gennes}, still remains one of hot subjects in modern statistical physics of polymers and wetting phenomena \cite{kawaguchi, bonn}. Such undiminished attention to the adsorption and localization of polymer chains at surfaces and interfaces is twofold. On one hand the interest is dealt with establishing the fundamental connection between polymer statistics and a general theory of phase transitions in condensed matter physics. On the other hand it is fueled by needs of applied polymer chemistry and design of new tailor-made materials. 

Here we focus on the adsorption problem of a pre-stretched polymer of length $L$ near an impermeable disc of radius $R$ with a sticky boundary. To demonstrate the universality of the phenomenon, we do not assume any specific microscopic mechanism of chain flexibility. Instead, we remain within the framework of an ideal two- or three-dimensional random walk, driven into a large-deviation fluctuation regime by imposing the constraint $L = N a = c R$.

Due to stickiness, the trajectories experience the localization transition at a critical boundary weight $\beta_{c} = e^{-u_c/(k_B T)}$. It is well-known that the partition function $\Omega_N$ of a $N$-step random walk touching the sticky boundary it is convenient to write as a product of partition functions $Z_{t_j}$ from $s$ independent "bridges" $t_1, t_2,..., t_s$ between  sequential arrivals at the boundary under the conservation condition $\sum_{j=1}^s t_j = N$ and summation over all possible $s$ ($1\le s<N$) -- see for example \cite{grosberg-kh, borisov1, borisov2}. Within each bridge the chain does not touch the boundary. Thus, the behavior of $\Omega_N$ is sensitive to the analytic structure of the "first return" partition function, $Z_{t}$, which can be written for stretched paths in the generic form as in \eq{eq:31}:
\be
Z_t \approx \mu^t e^{-\nu t^{1/3}}, \quad (\mu>0,\; \nu>0)
\label{eq:01}
\ee
where $\mu$ accounts for the statistical weight of each monomer in the bridge of a dimensionless length $t$ and is a non-universal quantity since it depends on the space dimensionality, lattice structure, etc, while the second "probabilistic" term reflects the nonalgebraic behavior of the return probability to the sticky boundary. Representing the polymer partition function $\Omega_N(\beta)$ as a collection of Brownian bridges, we get:
\be
\Omega_N(\beta) = \beta \sum_{s=1}^{\infty}\sum_{\{t_1+...+t_s=N\}} \prod_{j=1}^s (\beta Z_{t_j}) = \beta \sum_{s=1}^{\infty}\; \sum_{t_1=1}^{\infty}...\sum_{t_s=1}^{\infty}\Delta(t_1+...+t_s-N) \prod_{j=1}^s (\beta Z_{t_j})
\label{eq:02}
\ee
Using the integral representation of the Kronecker $\Delta$-function
\be
\Delta(x) = \frac{1}{2\pi i}\oint \frac{d\xi}{\xi^{x+1}}=\begin{cases} 1 & x=0 \\ 0 & x\neq 0 \end{cases}
\label{eq:03}
\ee
with $x=N-(t_1+...+t_s)$, we rewrite \eq{eq:02} as follows:
\be
\Omega_N(\beta) = \frac{\beta}{2\pi i}\oint \frac{d\xi}{\xi^{N+1}} \sum_{s=1}^{\infty}\left(\beta\sum_{t=1}^{\infty}(\mu\xi)^{t}e^{-t^{1/3}}\right)^s = \frac{\beta}{2\pi i}\oint \frac{d\xi}{\xi^{N+1}} \frac{1}{1-\beta\, \Theta(\xi)}
\label{eq:04}
\ee
where $\Theta(\xi) = \sum_{t=1}^{\infty}(\mu\xi)^{t}e^{-\nu t^{1/3}}$. Since the combination $\mu\xi=1$ corresponds to $N\to\infty$, the divergence of the denominator in \eq{eq:04} at $\xi=\mu^{-1}$ signifies the emergence of a phase transition at some value $\beta=\beta_{c}$ which is the solution of the equation $\beta_c \Theta(\xi=\mu^{-1})=1$. To find the dependence $\beta_c(\mu)$ we approximate the sum by the integral and obtain:
\be
\beta_{c} = \Theta^{-1}(\xi=\mu^{-1}) = \left(\sum_{t=1}^{\infty}e^{-\nu t^{1/3}}\right)^{-1} dt \approx \left(\int_0^{\infty} e^{-\nu t^{1/3}}\right)^{-1} = \frac{\nu^3}{6}
\label{eq:05}
\ee
The order of the phase transition, $\theta$, can be found by evaluating the first non-vanishing finite-size correction to the free energy, $F= -k_BT\ln \Omega_N(\beta)$, of the system at large but finite $N$:
\be
F(\beta) = F(\beta_c) + {\rm const}\,N |\beta-\beta_c|^{\theta} \qquad (|\beta-\beta_c|\ll 1)
\label{eq:06}
\ee
The value of $\theta$ is determined by the analytic behavior of the "loop factor" (i.e. on the first return function $Z_N$) which in our case in nonalgebraic. Expanding $\Theta(\xi)$ at $\xi\mu\nearrow 1$, i.e. taking $\xi$ in the form $\xi=\mu^{-1}-\delta$ where $0<\delta\ll 1$, we get the following expression 
\begin{multline}
\Theta(\xi=\mu^{-1}-\delta) \approx \int\limits_0^{\infty} e^{-(\mu \delta)t -\nu t^{1/3}} dt = \\ \left.\frac{\, _1F_2\left(1;\frac{1}{3},\frac{2}{3};\frac{\nu^3}{27 \delta}\right)}{\delta}-\frac{2 \pi  \nu  \text{Bi}\left(-\frac{\nu}{(-3\delta)^{1/3}}\right)}{(-3\delta)^{4/3}}\right|_{\delta\to 0} \approx \frac{6}{\nu^3} + \frac{360}{\nu^6}\mu \delta
\label{eq:07a}
\end{multline}
where $_1F_2(...)$ and $\text{Bi}(...)$ are correspondingly the hypergeometric and the Airy functions. Computing the series expansion of $\Theta^{-1}(\xi)$ up to the first leading term, we get
\be
\beta \approx \beta_c + 10 \mu |\delta| \qquad (|\delta|\ll 1)
\label{eq:07}
\ee
Note that the transition point in a chain of a finite length ($0<\xi<1$) has always bigger $\beta$ than the transition point $\beta_c=\nu^3/6$ in a chain of infinite length. According to \eq{eq:04}, the partition function in the localized phase at $N\gg 1$ is determined by the pole, $\xi_0$ of the function $\left(1-\beta\, \Theta(\xi)\right)^{-1}$ in the vicinity of the point $\xi=\mu^{-1}$. 

Defining the free energy per one monomer, $f(\beta)=F(\beta)/N$ at $N\gg 1$, we have:
\be
\Omega_N(\beta)\Big|_{N\gg 1} \approx (\mu \xi_0)^N = (1-\mu \delta)^N ; \qquad f(\beta) = \mu \delta
\label{eq:09}
\ee
Expressing $\delta$ from \eq{eq:07}, we get:
\be
\delta = \frac{|\beta-\beta_c|}{10\mu}
\label{eq:10}
\ee
Eq. \eq{eq:10} provides the dependence of $\delta$ on $|\beta-\beta_c|$ and hence determines the order of the phase transition $\theta$ (see \eq{eq:06}). Substituting \eq{eq:10} into \eq{eq:09} we get for $|\beta-\beta_c|\ll 1$:
\be
f(\beta) = \frac{\left|\beta-\beta_c\right|}{10} ; \quad \beta_c=\frac{\nu^3}{6}
\label{eq:11}
\ee
which means that the adsorption of the pre-stretched random path above a sticky disc happens as a first order phase transition.

\section{Discussion}
\label{sect:5}

We have investigated a special class of random walk problems in which the trajectory of the walker is strongly stretched in the vicinity of a convex impermeable surface---either a disc in two dimensions, or a sphere in three dimensions. While the standard return probabilities for random walks decay algebraically, we have demonstrated a fundamentally different behavior: a non-algebraic, stretched exponential decay of the first-return probability. This effect arises purely from geometric constraints and is shown to have significant implications, particularly for the adsorption behavior of polymers under an external stretching.

The physical setup involves a random walk consisting of $N$ steps of unit length $a$, whose total contour length $L = Na$ is scaled proportionally to the radius $R$ of the convex boundary such that $L = cR$, with $c > \pi$. This situation corresponds to an extreme stretching of the polymer that forces it to stay near the boundary. Using a combination of the optimal fluctuation approach and extensive numerical simulations, we have argued that the first-return probability $P_N$ up to the exponential accuracy is $P_N \sim \exp\left(-\mathrm{const} N^{1/3}\right)$.

This stretched exponential decay is reminiscent of the survival probability in disordered systems such as the classical Balagurov-Vaks trapping problem. In that case, the stretched exponential behavior arises due to spontaneous emergence of rare, trap-free regions in a disordered medium. In contrast, the present system is entirely free of disorder, and the non-algebraic decay is instead attributed to large-deviation effects induced by the interplay of curvature and stretching. 

We have confirmed the estimates based on the optimal fluctuation approach using path counting in two dimensions. The simulations fit the form $\exp(-\mathrm{const} N^{1/3})$ with a high confidence, demonstrating a very good agreement with the theoretical prediction.

Extending this analysis, we have analyzed a slightly different but related problem: a random loop pre-stretched in a proximity of an impermeable disc in two dimensions. The inflation of a loop is enforced by a magnetic field, such that the average area is fixed at $\pi R_c^2 = \pi (cR)^2$, with $c \geq 1$. The problem then reduces to a Schrödinger-type equation with a magnetic vector potential in polar coordinates, subject to Dirichlet boundary conditions at the boundary of a disc. Solving this equation in the vicinity of a disc boundary, we have obtained the asymptotic expression for the return probability of the pre-stretched loop in the form $P_N \sim \exp\left( -\frac{N}{4c^2} - \frac{|a_1|N^{1/3}}{4(\pi c)^{2/3}}\right)$, where $|a_1|$ is the first zero of the Airy function. The emergence in $P_N$ of the exponential term linearly proportional to $N$ reflects the different mechanism of stretching: here, the stretching is enforced a priori via external constraint (magnetic field) rather than arising self-consistently from a double-scaling limit. Nevertheless, the consistency in scaling structure with the earlier optimal fluctuation approach highlights the robustness of the non-algebraic behavior in such confined systems.

The last part of the paper explores how this non-algebraic return probability impacts the adsorption of elongated polymers. Specifically, we consider the adsorption of pre-stretched random walks near a sticky, impermeable disc in 2D. The full partition function for the polymer is constructed as a sum over sequences of bridges---parts of chains between successive contacts with the sticky boundary---each weighted by a first-return partition function of the form $Z(t) \sim \mu^t e^{-\nu t^{1/3}}$. Such a non-standard weight changes the analytic structure of the partition function and alters the nature of the adsorption transition. Whereas typical polymer adsorption (e.g., flexible chains in contact with a flat surface) undergoes a second-order transition, here we find that in the case of pre-stretched chains interacting with a sticky disc, the free energy exhibits a discontinuity, leading to a linear behavior $f(\beta) \sim |\beta - \beta_c|$, indicating a first-order phase transition. This finding aligns with known behavior in other systems: for instance, coil-to-globule transitions in rigid polymers are also first-order, in contrast to the continuous transitions seen in flexible chains \cite{grosberg-kh, birstein-pr, nouguchi}. Let us emphasize that the emergence of the first-order localization transition in our model is not a consequence of any specific mechanism of chain flexibility, but rather has a purely geometric–probabilistic origin: the random walk is driven into a large-deviation regime where the interaction with curved boundaries gives rise to atypical critical behavior.

Summarizing, our paper presents a unified framework where geometric constraints and external stretching, even in the absence of any disorder, can lead to atypical statistic behavior. Our work demonstrates that specific geometry and large deviations together could produce non-algebraic scaling laws in systems that would otherwise be expected to behave in a standard "algebraic" manner. 

\begin{acknowledgments}

Authors are grateful to A. Gorsky, D. Grebenkov, and A. Valov for fruitful discussions. S.N. acknowledges the hospitality of Beijing Institute for Mathematical Sciences and Applications (BIMSA) where the paper was finalized.

\end{acknowledgments}

\begin{appendix} 
\section{crossings test and first passage in the angle-based frame}
\label{sect:app}

The very concept of stretching becomes transparent via the following demonstration in two dimensions. Consider a stretched lattice random walk with endpoints located in the vicinity of two opposite poles ($\theta=0, r=R+\delta$) and ($\theta=\pi, r=R+\delta)$ of the disc, where $\theta$ is the polar angle and $\delta/R\ll 1$. Let the walk cross a radial wall at $\theta_0=\pi/2$ at time $t_0+\tau$, conditioned that it never crossed $\theta_0$ within the open interval $(t_0,t_0+\tau)$. Since this is an auxiliary problem that elucidates what stretching is, we present only the result of a numerical test without analytical derivations.

Specifically, simulations are carried out as follows. Take a two-dimensional square lattice with an impermeable disc of radius $R$ centered at $(R,0)$ and consider $N$-step bridges connecting $A=(-R,0)$ to targets $X=(R,R+\Delta y)$ lying above the apex and avoiding the disc interior. We vary the stretching parameter $c=N/R$ and, for each bridge $\mathcal{B}$, and count the number of crossings of the upper symmetry meridian, $K(\mathcal{B})|\{(x,y)\in \mathbb{Z}^2:\; x=0,\; y\ge R\}$. We are interested in the average value of $\langle K\rangle$ over all reachable bridges for a fixed $c$. By "all paths for fixed $c$" we mean that $N$ and $R$ are related by $N=cR$, and the end point $X$ is allowed to slide along a vertical ray over the vertex. Thus, for a given $c$ we traverse every $\Delta y\ge 0$ that is reachable by at least one $N$-step bridge from $A$ to $X$. For each offset, we enumerate all reachable bridges, and for every individual path we count the number of crossings of the symmetry meridian. 

In a semi–log plot the average number of returns $\langle K\rangle$ at a point $X$ in a stretched regime $La = c R$, ($4\le c \le 12$) decreases nearly linearly with decreasing $c$:
$$
\langle K\rangle \sim K_0\, e^{q |c-c_0|}.
$$
where $q \approx 0.074 \pm 0.001$ and $c_0=4$. Thus, in the angle-based frame, the number of returns is exponentially suppressed as $c$ tends to its lower lattice boundary value, $c_0$.

\begin{figure}[ht]
  \centering
  \includegraphics[width=0.62\textwidth]{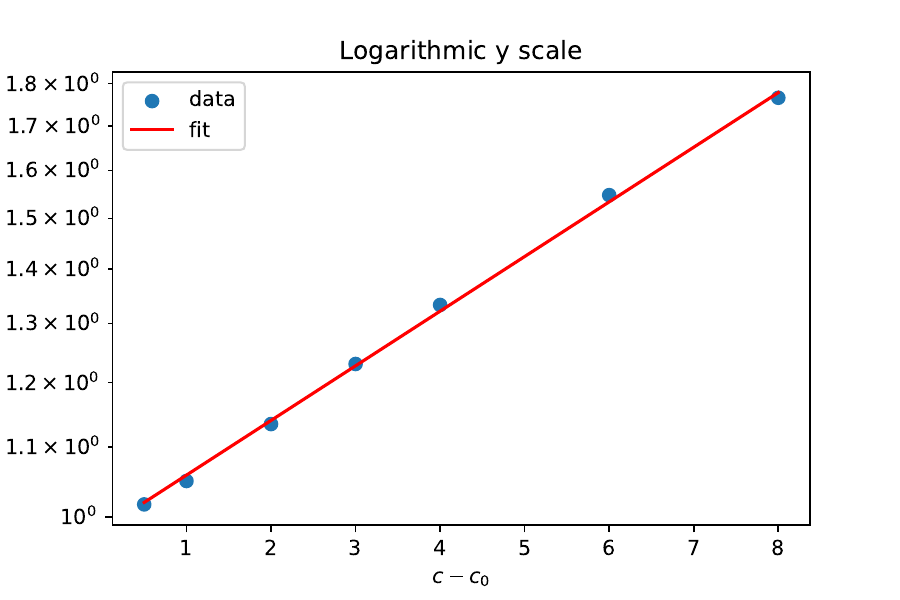}
  \caption{Mean number of meridian crossings $\langle K\rangle$ in a semi-logarithmic scale as function of relative stretching $c-c_0$, where $c_0=4$.}
  \label{fig:S1}
\end{figure}
\end{appendix}

\bibliography{ref}  

\end{document}